\documentclass[%
 aip,
 amsmath,amssymb,
 reprint,%
]{revtex4-2}

\usepackage{graphicx}
\usepackage{dcolumn}
\usepackage{bm}

\usepackage[utf8]{inputenc}
\usepackage[T1]{fontenc}
\usepackage{mathptmx}
\usepackage{mhchem}
\usepackage{here}
\usepackage{float}
\usepackage{soul}
\usepackage[dvipsnames]{xcolor}
\usepackage{ulem}

\begin{document}

\preprint{AIP/123-QED}

\title[]{ 
 Effect of boron nitride defects and charge inhomogeneity on 1/f noise in encapsulated graphene  }

\author{Chandan Kumar}
\email{kchandan@iisc.ac.in}
 \altaffiliation[Present address: ]{Department of Condensed Matter Physics, Weizmann Institute of Science, Israel.}
\author{Anindya Das}%
 
\affiliation{ 
Department of Physics, Indian Institute of Science, Bangalore 560 012, India
}%

\date{\today}

\begin{abstract}
Low frequency $1/f$ noise is investigated in graphene, encapsulated between hexagonal boron nitride (hBN) substrate in dual gated geometry. The overall noise magnitude is smaller as compared to graphene on \ce{Si/SiO2} substrate. The noise amplitude in the hole doped region is independent of carrier density while in the electron doped region, a pronounced peak is observed, at  Fermi energy, $E_F \sim 90$ meV.  The physical mechanism of the anomalous noise peak in the electron doped region is attributed to the impurity states originating from the Carbon atom replacing the Nitrogen site in hBN crystal. Furthermore, the noise study near Dirac point shows characteristic ``M-shape'',  which is found to be strongly correlated with the charge inhomogeneity region near Dirac point. 
\end{abstract}

\maketitle

Hexagonal boron nitride (hBN) has evolved as one of the most fundamental ingredient in the fabrication of van der Waals heterostructures. 
The clean and flat surface of hBN has led to the tremendous improvement in graphene device quality. This has led to the observation of intrinsic properties of graphene, which were obscured previously\cite{gorbachev2012strong,taychatanapat2013electrically,bolotin2009observation,young2012spin,
amet2014selective,kumar2018equilibration}.
Furthermore, due to the same lattice structure as that of graphene, hBN-graphene heterostructure creates 
a periodic super-lattice potential\cite{yankowitz2012emergence}. This has opened a new field of moire physics\cite{ponomarenko2013cloning,dean2013hofstadter,barrier2020long}. Moreover, hBN is being used as an encapsulating layer in various 2D materials such as semiconductor transition metal
dichalcogenides (TMDs), superconductor: \ce{NBSe2}, semimetal: black phosphorous, ferromagnet: \ce{CRI3} \cite{huang2017layer,khestanova2018unusual,cao2015quality,jin2019observation}. Encapsulating these materials with hBN leads to reduction in charge inhomogeneity and also protects the material from degradation. 
Using hBN as a dielectric in all these material it is assumed that it will not degrade the intrinsic property of the 2D material.

Recently, it has been reported that the hBN crystal gets unintentionally doped with Carbon during the growth process\cite{onodera2019carbon,watanabe2019far,onodera2020carbon}. 
Moreover, the carbon rich domain can exist even after the exfoliation process and thus can hamper the device performance. Since, hBN act as a base substrate, tunnel barrier or as an encapsulating layer in van der Waals materials, it becomes prerequisite to distinguish the Carbon rich hBN from pure hBN crystal and understand the influence of Carbon rich hBN on the 2D material.

We have used $1/f$ noise as an ultra-sensitive spectroscopic tool to probe the impurity state in hBN and study the influence of impurity level of hBN on mono layer graphene. $1/f$ noise has been studied extensively in graphene on \ce{Si/SiO2} substrate in last several years revealing very different noise trend\cite{xu2010effect,rumyantsev2010electrical,heller2010charge,zhang2011mobility,
pal2011microscopic,kaverzin2012impurities,lin2008strong,pal2009ultralow} unlike conventional semiconductors which follows Hooge's emperical relation. The deviation from the Hooge's relation is mainly explained in terms of electron and hole puddles present near the Dirac point\cite{xu2010effect,pellegrini20131}, inhomogeneous charge trap distribution\cite{kaverzin2012impurities,pellegrini20131}, interplay between short range and long range scattering\cite{kaverzin2012impurities,zhang2011mobility} and the effect of contacts\cite{shao2009flicker,heller2010charge,karnatak2016current}.
The noise study of graphene on hBN substrate shows drastic reduction in noise value as compared to graphene on \ce{Si/SiO2} substrate
and noise away from the Dirac point is independent of carrier density\cite{stolyarov2015suppression,kayyalha2015observation,kumar2016tunability,kakkar2020optimal}. However, there is no consensus on the noise behaviour close to Dirac point\cite{stolyarov2015suppression,kayyalha2015observation,kumar2016tunability,kakkar2020optimal}. Furthermore, noise study in graphene using Carbon rich hBN as a substrate is still lacking, there is no report of noise study in any 2D material on Carbon rich hBN substrate.

In this work, we have carried out a detailed $1/f$ noise study in hBN encapsulated graphene in dual gated geometry.
We find that the noise amplitude in the hole doped regime is  independent of carrier density ($n> 1.7 \times 10^{11} cm^{-2}$). Contrary to this, in the electron doped regime, we observe a strong peak in noise amplitude around $E_F \sim 90$ meV, although we don't observe any signature of resistance increase around this Fermi energy. The noise amplitude peak is found to be associated  with the impurity states originating from the Carbon doping at the Nitrogen sites in hBN crystal. Furthermore, noise near the Dirac point shows characteristic ``M-shape'', which is found to be associated with the charge inhomogeneity near the Dirac point.

\begin{center}
	\begin{figure*}[htp!]
		\includegraphics[width=1\textwidth]{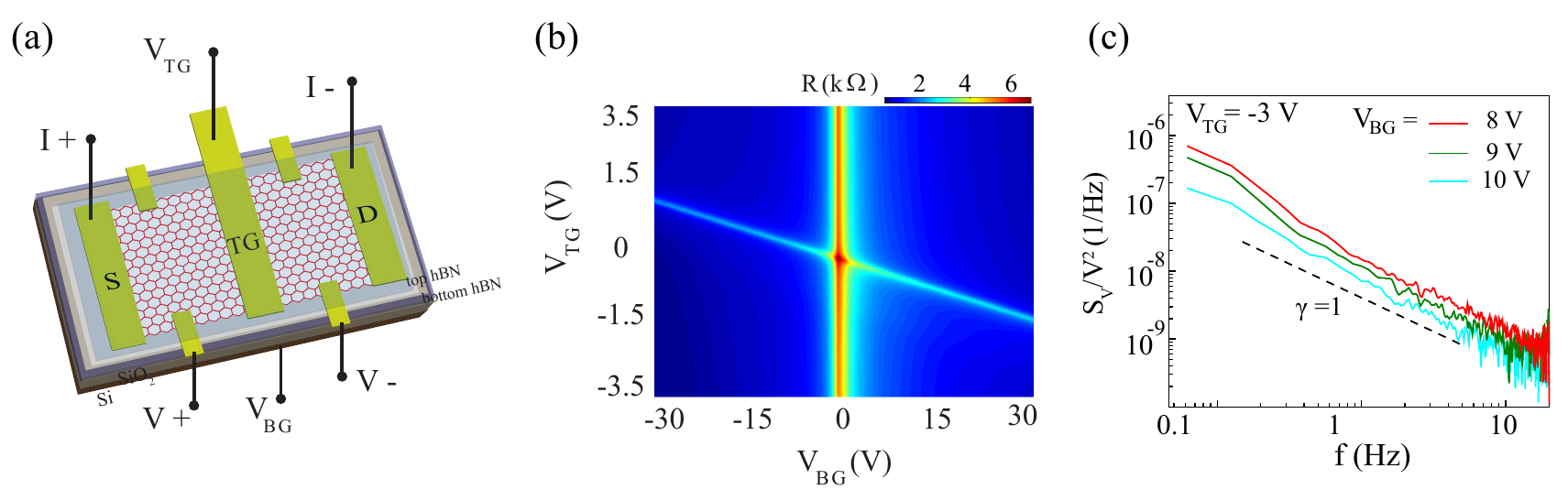}
		\caption{ (a) Schematic of the measured graphene device. The graphene is encapsulated between the top and bottom hBN. The 285 nm thick \ce{SiO2} acts as a back gate dielectric and degenerately doped silicon as the back gate. The back gate controls the density throughout the device while top gate which is defined on a small portion of top hBN controls density only on a small segment of graphene. (b) 2D color map of resistance as a function of $V_{BG}$ and $V_{TG}$ measured in four probe geometry at 77 K. The sharp vertical and diagonal line represents the Dirac point in the back gate and top gate region, respectively.  From the slope of diagonal line we calculate the top hBN thickness $\sim 11 $ nm. (c) Normalized noise spectral density as a function of frequency at $V_{TG} =$ -3 V for different back gate voltages.}
		\label{fig:images}
	\end{figure*}
\end{center}

Our device is a stack of hBN/Graphene/hBN. It is prepared following the dry transfer technique\cite{zomer2011transfer}. First a thick hBN ($\sim 30$ nm) is exfoliated on \ce{Si/SiO2} substrate. A piece of glass slide is prepared with a layer of pmma and graphene is exfoliated on the pmma. The glass slide containing the graphene is then loaded on micro manipulator and the  graphene flake is transferred on the hBN. The prepared stack of hBN/Graphene on \ce{Si/SiO2} is then cleaned in acetone followed by IPA cleaning. Standard lithography technique is used to design the contacts which is followed by thermal evaporation of Cr (5 nm)/Au (70 nm) at a base pressure of 3e-7 mbar. The prepared device is then vacuum annealed at 300 $^\circ$C for 3 hours. This is followed by transfer of a thin hBN ($\sim 10$ nm) on prepared stack of hBN/Graphene. This thin top hBN act as a dielectric and top gate is defined using another lithography step followed by thermal evaporation of Cr (5 nm)/Au (70 nm).

Figure 1(a) shows the device schematic. The degenerately doped silicon (Si) acts as a global back gate ($V_{BG}$) with 285 nm thick silicon dioxide (\ce{SiO2}) as the dielectric. The global back gate controls the carrier density throughout the graphene channel while the local top gate ($V_{TG}$) tunes the density on a very small region of graphene (Fig. 1(a)). The length and width of graphene channel is 8 $\mu m$ and 2 $\mu m$, respectively. The distance between voltage probe is 4 $\mu m$ and top gate width is  1 $\mu m$. Figure 1(b) shows the 2D color plot of four probe resistance as a function of $V_{BG}$ and $V_{TG}$. The diagonal and vertical lines represents the Dirac point under the top gate and back gate region, respectively. From slope of the diagonal line, we calculate the top hBN thickness $\sim$ 11 nm.  All the measurements are performed in liquid Nitrogen at 77 K.

The $1/f$ noise is generally characterized using  Hooge's emperical relation:
\begin{equation}
S_R/R^2 = S_V/V^2 = S_I/I^2 = \alpha_H / Nf^{\gamma}.
\end{equation} 
where $S_I/I^2$ is the normalized current noise density, $f$ is the frequency, $I$ is the current, $N$ is the total number of charge carriers and $\alpha_H $ is the Hooge parameter. The exponent $\gamma$ is ideally expected to be 1 for $1/f$ noise. The noise measurement is performed following the measurement scheme discussed in our previous work\cite{kumar2016tunability,kumar2018equilibrationssc}. 

The noise measurements were performed at a constant current bias of 200 nA.
Figure 1(c) shows the normalized current noise density ($S_V/V^2$) at $V_{TG} = -3$ $V$ for three representative $V_{BG}$. We see clear $1/f$ behaviour. The black dash line shows the pure $1/f$ noise behaviour with $\gamma=1$.  For our measurement we find $\gamma$ in between 0.9 to 1.1 at various carrier concentration, suggesting that the charge traps are distributed randomly.\cite{balandin2002noise,stolyarov2015suppression}\\
The noise amplitude (A) is defined as:\cite{balandin2013low}
\begin{equation}
A = \frac{1}{N} \sum_{m=1}^{N} f_m S_{Im}/I_m^2 
\end{equation} 

Here $S_{Im}$ and $I_m$ are the noise spectral density and source-drain current at m different frequencies $f_m$. This method of defining noise amplitude helps to reduce the measurement error which comes by measuring noise amplitude at particular frequency\cite{xu2010effect,lin2008strong}.

Figure 2 (a)-(c), plots the noise amplitude as a function of $V_{BG}$ at $V_{TG} =$ -3 V, 0 V and 3 V, respectively. 
The corresponding four probe resistance is presented in the bottom panel (Fig. 2 (d)-(f)). The resistance peak close to $V_{BG}= -6$ (Fig. 2(e)) and $V_{BG}= 0$ (Fig. 2 (d-f)) corresponds to the Dirac point in the top gate and back gate region, respectively. We find that the over all noise amplitude value is smaller compared to graphene on \ce{Si/SiO2} substrate. This is consistent with the previous reports\cite{stolyarov2015suppression,kayyalha2015observation,kumar2016tunability,kakkar2020optimal} . More interestingly, we find a strong noise amplitude peak far away from the Dirac point, in the electron doped region at $V_{BG} \sim 8$ V $\sim E_F = 90$ meV $(n \sim 6.4 \times 10^{11} cm^{-2})$, although the resistance curves do not show any special feature around this density. We also find that this noise peak is independent of the top gate voltages. On the contrary, for the hole doped region ($V_{BG}<-3$ in Fig. 2 (a)-(c) and $V_{BG}<-10$ in Fig. 2(b)) the noise is independent of carrier density. The noise behaviour close to Dirac point is very intricate and we shall discuss it in coming section.

\begin{center}
	\begin{figure*}[ht!]
		\includegraphics[width=.8\textwidth]{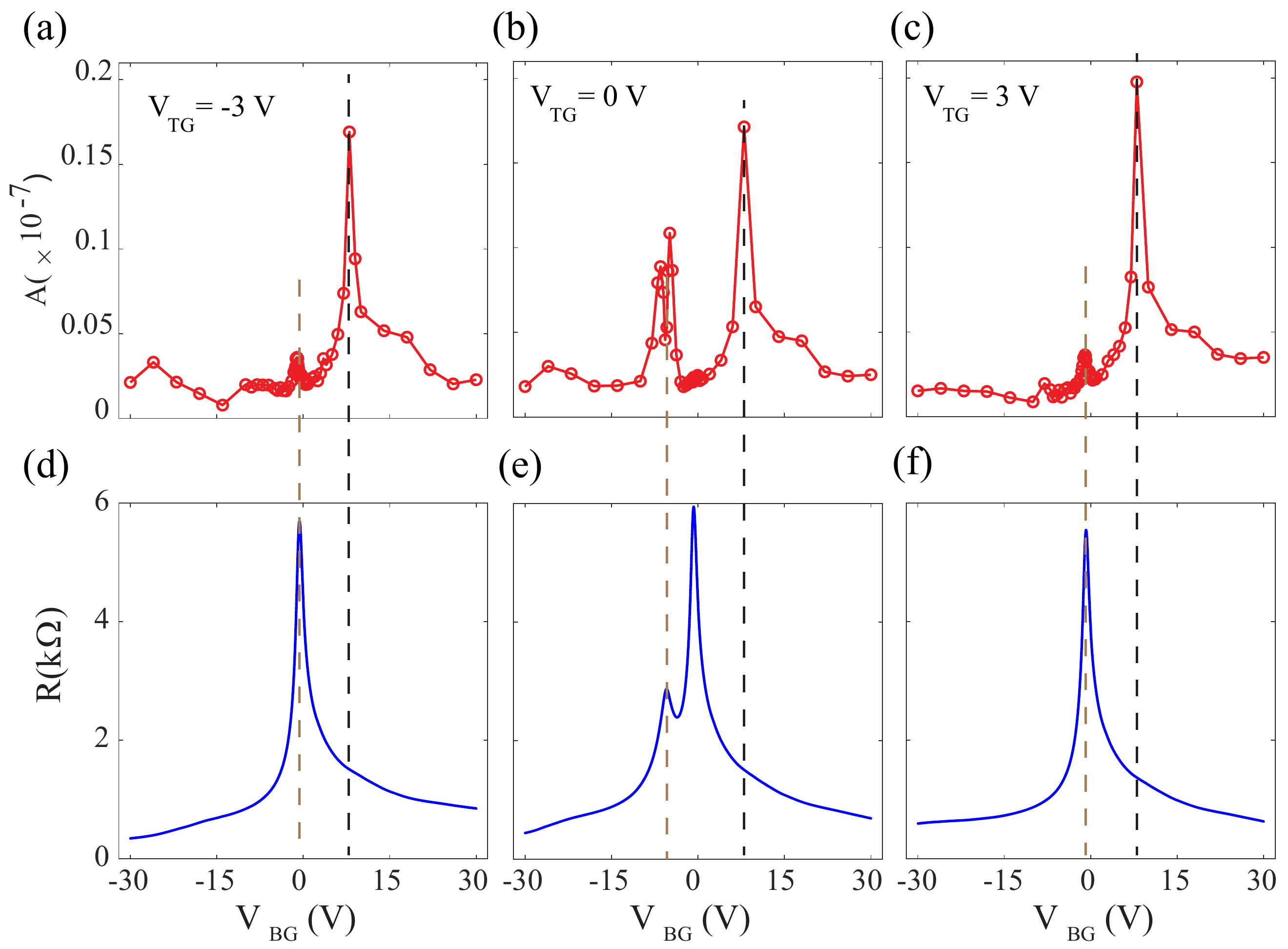}
		\caption{ Noise amplitude (top panel) and the corresponding four probe resistance (bottom panel) as a function of back gate voltage at three different top gate voltages, $V_{TG} =$ -3 V, 0 V and 3 V.. The resistance curves are 		obtained from Fig. 1(b) by taking horizontal line slice at $V_{TG} =$ -3 V, 0 V and 3 V.
		Overall noise amplitude is smaller as compared to graphene on \ce{Si/SiO2} substrate. Along with a small noise increment near the Dirac point, a sharp noise peak is observed in 			the electron doped region at 
		$V_{BG} \sim$ 8 V ( $n$ $\sim$ 6.4 $\times 10^{11}$ $cm^{-2}$. In contrast, noise is independent of carrier density in the hole doped region.
		  The resistance peak in Fig. 2(e) near $V_{BG} \sim$ -6 V is due to the Dirac point in the top gate region (Diagonal line in Fig. 1(b)). The dashed black and brown lines mark the noise peak in the back gate region at $n$ $\sim$ 6.4 $\times 10^{11}$ $cm^{-2}$ and the Dirac point in top gate region, respectively.}
		\label{fig:images}
	\end{figure*}
\end{center}

In this section, we explore the origin of noise peak at $V_{BG} \sim$ 8 V ( $n$ $\sim$ 6.4 $\times 10^{11}$ $cm^{-2}$ and $E_F \sim$ 90 meV). 
Recent study of Onodera \textit{et al.}\cite{onodera2019carbon,onodera2020carbon} in graphene on Carbon doped hBN substrate show that carbon rich domain can exist even after mechanical exfoliation process and it affects the magneto-resistance drastically. Since the doped hBN flakes can’t be differentiated from the pristine hBN flakes in an optical microscope, photo luminance technique was used to identify the hBN flakes with carbon doped region and the pristine region.  The graphene devices on pristine hBN and carbon doped hBN resulted in devices with ultra-high mobility and poor mobility, respectively. Furthermore, using  magneto-resistance study, the energy of Carbon impurity level was estimated around  $\sim 80 - 110$ meV. The DFT calculation in AB-stacked hBN-graphene heterostructure showed the presence of addition impurity states at $ \sim$ 150 meV and $\sim$ 1.35 eV above the Dirac point for the Nitrogen and Boron site being doped by Carbon in hBN crystal, establishing that the magneto-resistance anomaly observed in graphene is due to Carbon doping at Nitrogen sites in hBN crystal (Fig. 3(a))\cite{onodera2020carbon}.We plot in Fig 3(b) the impurity states measured by  Onodera \textit{et al.} \cite{onodera2020carbon}, due to Carbon doping at Nitrogen site in the hBN crystal. We find that the Fermi energy of the observed noise peak ($\sim$ 90 meV) in Fig. 2 (a)-(c) 
lies in the impurity energy range (80-110 meV), obtained by Onodera \textit{et al.} \cite{onodera2020carbon},
indicating that the noise peak in the electron doped regime is associated with the acceptor like impurity states formed due to Carbon doping at Nitrogen sites in hBN crystal. Schematic of the Carbon defects in the hBN layer is shown in Fig. 3(a). Defects states are distributed randomly at different heights from the adjacent graphene layer. The Carbon atom substituted at the Nitrogen sites in hBN layer is presented in Fig. 3(a) zoom in.

\begin{center}
	\begin{figure*}[ht!]
		\includegraphics[width=1\textwidth]{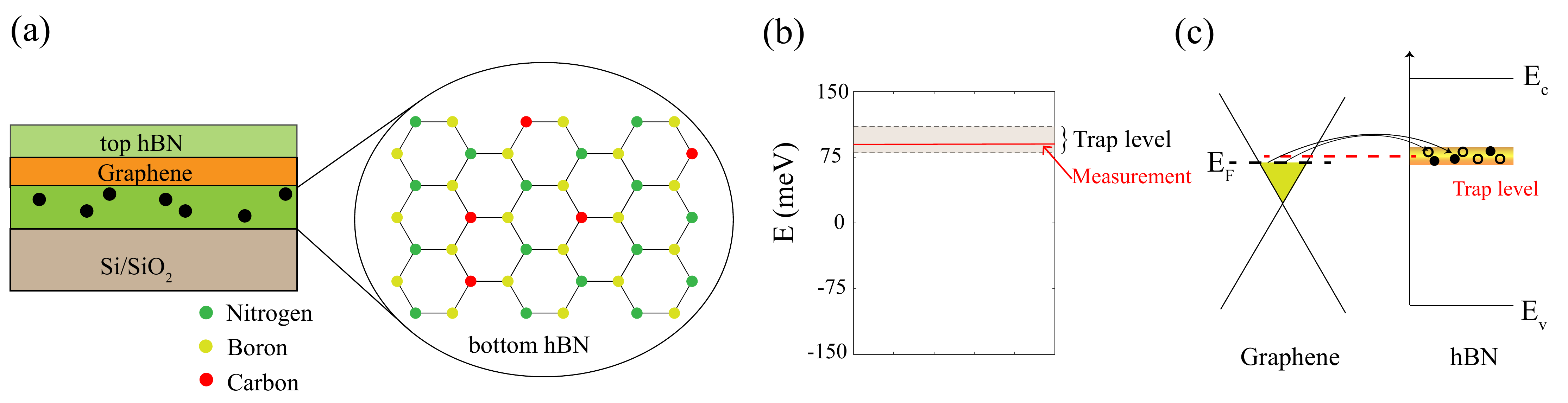}
		\caption{(a) Illustration of Carbon defects in the bottom hBN (black dots). These defects states are at different height from the graphene layer. Zoom in on a single sheet of hBN layer and showing doping by Carbon atoms at the Nitrogen sites in hBN crystal. (b) The horizontal stripe represents defects energy states of Carbon doping the Nitrogen sites(CN) in hBN crystal, obtained from ref \citep{onodera2020carbon}. The Fermi energy of the anomalous noise peak observed in our experiment is $\sim 90$ meV (red horizontal line).The energy levels are with respect to Dirac point in graphene. (c) Illustration of charge trapping and de-trapping process by the impurity states (CN) in hBN when the Fermi level in graphene is close to the impurity states (CN) in hBN. The shaded region (orange), denotes the thermal broadening of impurity states in hBN.} 		
		\label{fig:images}
	\end{figure*}
\end{center}

\begin{center}
	\begin{figure*}[htp!]
		\includegraphics[width=1\textwidth]{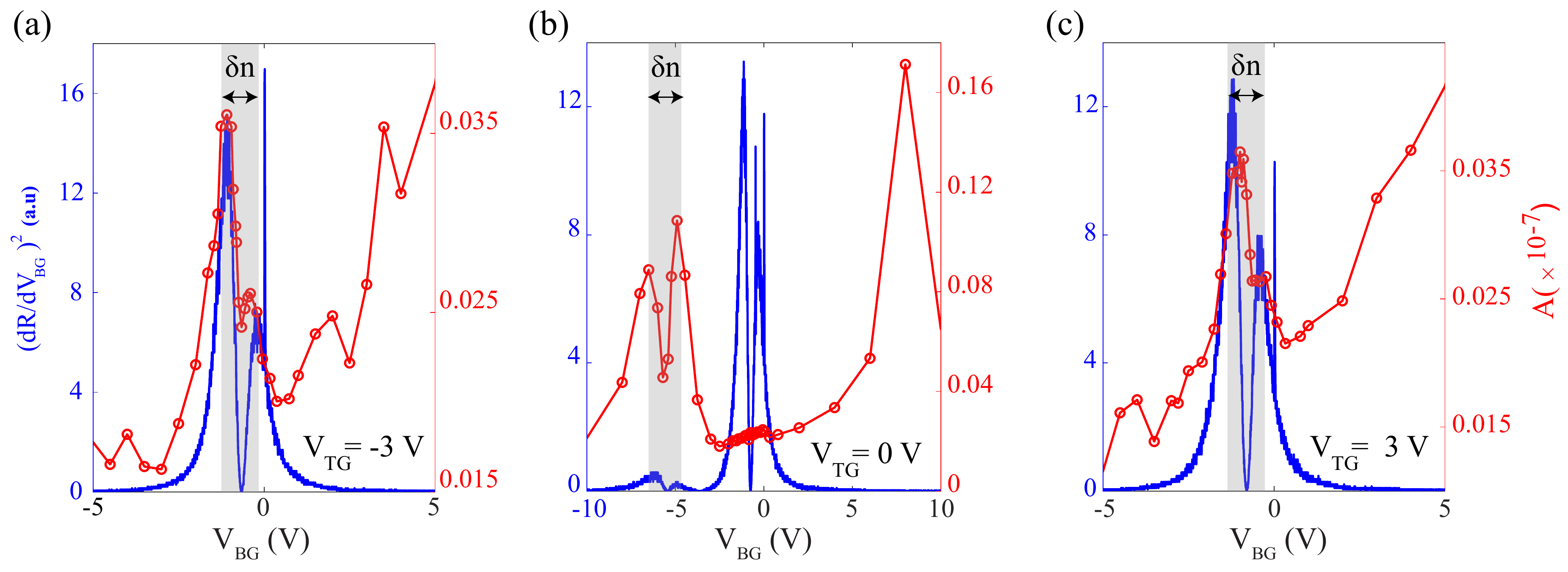}
		\caption{ Noise amplitude (right axis) and $(dR/dV_{BG})^2$ (left axis) as a function of back gate voltage near the Dirac point at three different top gate voltages, $V_{TG} =$ -3 V, 0 V and 3 V. The difference between peaks and dip in $(dR/dV_{BG})^2$ corresponds to charge inhomogeneity($\delta n \sim 5\times 10^{10} cm^{-2}$ and Dirac point, respectively. The black shaded stripe shows the charge inhomogeneity region of the device. The dip and peak in noise amplitude coincides with the dip and peak of $(dR/dV_{BG})^2$, establishing that noise is minimum at the Dirac point. In between the Dirac point and charge inhomogeneity, noise increases with its peak at the charge inhomogeneity. Doping the graphene beyond the charge inhomogeneity point leads to decrease in noise amplitude in the hole doped region.}
		\label{fig:images}
	\end{figure*}
\end{center}

The physical mechanism of $1/f$ noise in graphene has been explained using the Hooge's mobility fluctuation\cite{zahid2013reduction,rumyantsev2013effect,vail20201,cultrera2018role,takeshita2016anomalous} or McWhorter charge number fluctuation method\cite{peng2017carrier,wang2021effect} or both. 
Furthermore, both these mechanisms are correlated \cite{stolyarov2015suppression,vail20201}.
Assuming that hBN is free of trap states or dangling bonds, the low noise amplitude in graphene on hBN substrate were explained by screening of charge carriers in graphene from the trap states in \ce{Si/SiO2} by the hBN layer \cite{stolyarov2015suppression,kayyalha2015observation}.

Here we consider the effect of Carbon doped hBN on noise amplitude in the adjacent graphene layer. As discussed earlier, doping of Carbon atom at the Nitrogen sites in hBN crystal creates an impurity states $\sim 80-110$ meV above the Dirac point in the hBN (Fig. 3(b)). These defect states can act as a strong scattering center leading to trapping / de-trapping of charge carrier in graphene implying charge number fluctuation in the graphene layer.  The trapping/ de-trapping events (Fig. 3(c)) will be maximum when the Fermi level in graphene layer matches with the defect energy level in hBN, leading to large noise peak. As we move away from the defect energy level, the trapping/de-trapping events will reduce leading to decrease in noise amplitude. Our work is in very good agreement with the recent theoretical work by Francesco $et. al$ \cite{pellegrino20201}, where they predicted noise peak in the conduction band of graphene, assuming McWhorter charge number fluctuation model. The noise peak was associated to the trap energy centred far away form the Dirac point in the conduction band.

We now focus on the noise behaviour close to Dirac point.  Figure 4 shows the noise amplitude (right Y axis) and $(dR/dV_{BG})^2 $ (left Y axis) as a function of back gate voltage at three different top gate voltages around the Dirac point. The dip and peaks in $(dR/dV_{BG})^2 $ corresponds to the Dirac point and the charge inhomogeneity ($\delta n$) of the device. We observe clean ``M-shape'' noise behaviour near the Dirac point in the top gate region (Fig. 4b). Furthermore,
the noise amplitude peaks and dip coincide with $(dR/dV_{BG})^2 $ peak and dips, suggesting that noise is minimum exactly at the Dirac point. Noise value increases in between the Dirac point and $\delta n$, with the noise peak at $\delta n$. With the further increase in carrier density ($ n > \delta n$), noise decreases and becomes almost flat beyond $n \sim 5 \times 10^{11} cm^{-2}$, for the hole doped region. Fig. 4a and Fig. 4c shows noise behaviour resembling "M-shape" with one peak having larger magnitude than the other peak and not so prominent dip at the Dirac point. This is referred as "asymmetric M-shape" noise, which may arise due to the intricate and convoluted way in which Fermi energy and trap energy vary over the device area\cite{pellegrini20131}. We also observe smaller noise amplitude in the back gate region, without "M-shape" (Fig. 4b). This may be related to the gate area size. 
Since back gate and top gate has different area we can compare the area normalized noise amplitude, $A_N=A\times L\times W$ where A, L and W are the noise amplitude, length and width of the top gated or back gated region.  We find very similar area normalized noise amplitude in top ($A_{N_{TG}}$), and back gate ($A_{N_{BG}}$) region with $A_{N_{TG}} \sim 1.78 \times 10^{-8}$ and $A_{N_{BG}} \sim 1.84 \times 10^{-8}$. The smearing of "M-shape" noise due to spatial variation of Fermi energy or charge inhomogeneity can lead to the disappearance of "M-shape" noise behaviour\cite{pellegrini20131}.

The charge inhomogeneity model of Xu \textit{et al.}\cite{xu2010effect}, which has been used in past to explain noise
in graphene on \ce{Si/SiO2} substrate can qualitatively explain the ``M-shape'' noise behaviour near Dirac point in our device. According to Hooge's relation, noise at Dirac point can be written as $A \sim \frac{2\alpha_H}{N} \sim \frac{2\alpha_H}{\delta n D}$. Here factor of 2 is because both electron and hole puddle contribute to the noise, $N=N_e\sim N_h$, $\delta n$ and D corresponds to the charge inhomogeneity and the electron-hole puddle size, respectively. Between the Dirac point and charge inhomogeneity point the noise amplitude can be written as $A =  \frac{\alpha_H}{n_{e} D_e} + \frac{\alpha_H}{n_{h} D_h}$. The application of positive gate voltage will increase electron density. Moreover, it will also increase electron puddle size and decrease the hole puddle size. Thus, in this region the noise contribution due to hole (minority carrier) is greater than the electrons (majority carrier) and hence noise increases. Beyond the spatial charge inhomogeneity region, the minority carrier puddle region (holes) starts shrinking and can no longer contribute to the conduction. Hence the majority carriers (i.e electrons) starts contributing to the noise and total noise is given by $A \sim A_e \sim \frac{\alpha_H}{N}$, implying decrease in noise value with increasing carrier density. Thus, the noise at the Dirac point is minimum, between the Dirac point and charge inhomogeneity point noise increases and is dominated by the minority charge carriers and beyond charge inhomogeneity point noise decreases and is governed by majority charge carriers. At higher hole density ($n> 5 \times 10^{11} cm^{-2}$) noise becomes independent of gate voltage, which is consistent with the literature\cite{stolyarov2015suppression,kayyalha2015observation}.

In conclusion, we show that $1/f$ noise is an extremely sensitive tool to probe the defects states in the hBN substrate. The defects states arising due to Carbon doping at the Nitrogen sites in hBN crystal manifests itself as noise peaks in the adjacent graphene layer, although no signature is visible in the transport measurement. Furthermore, we show that noise near the Dirac point is dominated by inhomogeneous charge distribution of the device. We believe, our study will establish $1/f$ noise as a effortless technique to distinguish Carbon rich hBN from pure hBN crystal.\\ 

Authors thank Francesco M. D. Pellegrino for enlightening discussions.

\vspace{1 cm}
\textbf{DATA AVAILABILITY}\\
The data that support the findings of this work are available from
the corresponding authors upon reasonable request.

\vspace{1cm}
\textbf{REFERENCES}
\bibliographystyle{apsrev4-1}
\bibliography{ref}


\end{document}